\documentclass[11pt]{article}
\usepackage{epsf}

\def\lsi{\raise0.3ex\hbox{$<$\kern-0.75em\raise-1.1ex\hbox{$\sim$}}}
\def\gsi{\raise0.3ex\hbox{$>$\kern-0.75em\raise-1.1ex\hbox{$\sim$}}}

\newcommand{\gsim}{\mathop{\gsi}}

\begin{document}
\begin{flushright}
BI-TH-97/24 \\
SU-4240-666\\
September 1997
\end{flushright}

\begin{center}
\vspace{24pt}

{\Large \bf Improved Algorithms for Simulating
 Crystalline Membranes}
\vspace{24pt}

{\large \sl Gudmar Thorleifsson$\,^1$ \,{\rm  and}\, 
            Marco Falcioni$\,^2$}
\vspace{6pt}

$^1$Fakult\"{a}t f\"{u}r Physik, Universit\"{a}t Bielefeld,
 D-33615 Bielefeld, Germany\\

\setcounter{footnote}{1}

$^2$Physics Department, Syracuse University,
 Syracuse NY 13244-1130, USA\footnote{Current Address: Chemical
 Engineering Department, University of California, Los Angeles, CA,
 90095-1592, USA}\\

\vspace{10pt}

\begin{abstract} 
The physics of crystalline membranes, i.e.\ fixed-connectivity
surfaces embedded in three dimensions and with an extrinsic curvature
term, is very rich and of great theoretical interest.  To understand
their behavior, numerical simulations are commonly used.
Unfortunately, traditional Monte Carlo algorithms suffer from very
long auto-correlations and critical slowing down in the more
interesting phases of the model.  In this paper we study the
performance of improved Monte Carlo algorithms for simulating
crystalline membrane, such as hybrid overrelaxation and unigrid
methods, and compare their performance to the more traditional
Metropolis algorithm.  We find that although the overrelaxation
algorithm does not reduce the critical slowing down, it gives an
overall gain of a factor 15 over the Metropolis algorithm.  The
unigrid algorithm does, on the other hand, reduce the critical slowing
down exponent to $z \approx 1.7$.
\end{abstract}
\end{center}

\vspace{15pt}

\section{Introduction}
When using Monte Carlo methods to study physical systems one is
usually faced with the problem of {\it critical slowing-down} (CSD) in
the critical region were a typical correlation length of the system
diverges.  That is, the auto-correlation time of traditional Monte
Carlo algorithms, the time it takes to generate ``statistically
independent'' configurations, grows rapidly with system size.  For
example, for the well known Metropolis and heat-bath algorithms this
time grows linearly with the system size, making simulations of large
systems prohibitively time consuming.

It is thus very important to construct new Monte Carlo algorithms that
can reduce CSD.  Much progress has been made in this direction;
examples of improved algorithms are, to name a few: Adler's
overrelaxation \cite{adler}, Fourier acceleration \cite{lan},
multigrid \cite{orgmult} and cluster algorithms \cite{swendsen}.
Those algorithms have been applied successfully to variety of models
and, in some special cases, have eliminated CSD altogether.  This,
though, is usually accomplished only for relatively simple models,
such as free field theories or spin models with simple interactions.
For more complicated models, where sophisticated methods are harder to
implement, the improvement is usually somewhat less.

In this paper we study the performance of two such improved
algorithms, overrelaxation and multigrid, in simulations of {\it
crystalline membranes}.  Crystalline membranes are internally rigid
surfaces, embedded in three dimensions, with an extrinsic curvature
term.  They exhibit a phase transition between a high-temperature
crumpled and a low-temperature flat phase.  It is especially in the
flat phase, and at the {\it crumpling} transition, that Monte Carlo
simulations with traditional algorithms suffer from very long
auto-correlations.

As with many interesting models, the Hamiltonian of a crystalline
membrane is too complicated for a direct implementation of the methods
we want to employ.  Some simplifications have to be made.  For the
overrelaxation we use a quadratic approximation to the Hamiltonian
when choosing a new trial position --- this requires an additional
accept-reject step to restore detail balance.  This is usually
referred to as {\it hybrid} overrelaxation. Instead of multigrid we
use a simpler implementation, the {\it unigrid} algorithm, in which a
coarsening transformation of the field configuration is not needed ---
the fields at the original (fine) lattice are simulated at all levels.

In addition to the performance of the algorithms, we also examine the
importance of randomness in the updating procedure, i.e.\ the order in
which the fields are updated.  According to standard folklore, too
much randomness in the updating procedure increases CSD as the system
takes a drunkard's walk through the phase space.  Too little noise, on
the other hand, increases CSD as well since the system is too weakly
ergodic.  It is thus important to tune the amount of randomness in the
algorithm appropriately.

The paper is organized as follows.  In Section 2 we describe the
particular model of a crystalline membrane we study, discuss the
problems of the simulations and the performance of the Metropolis
algorithm.  In Section 3 we describe the hybrid overrelaxation
algorithm and our approximation.  In Section 4 we test the performance
of the unigrid algorithm.  Finally, in the Section 5, we compare the
overall performance and merits of those different methods and comment
on possible further improvements and applications.


\section{A model of crystalline membranes}

The model we simulate is a simple discretization of a phantom (non
self-avoiding) crystalline membrane, inspired by the Polyakov action
for Euclidean strings with extrinsic curvature \cite{polyakov,jan}.  A
discrete crystalline membrane is described by a regular
two-dimensional triangulation embedded in three-dimensional space
where it is allowed to fluctuate.  The Hamiltonian is composed of two
terms: a pair potential between neighboring nodes and a bending energy
term.  As a pair potential we use a simple Gaussian spring potential,
and we model the bending energy by a ferromagnetic interaction between
neighboring normals to the faces of the surface:
\begin{equation}
\label{e21}
{\cal H} \;=\; \sum_{\langle ij \rangle} |\vec{r}_i - \vec{r}_j |^2
   \;+\; \kappa \sum_{\langle ab \rangle} (1 - \vec{n}_a \cdot \vec{n}_b ) .
\end{equation}
Here $i,j$ label the intrinsic position of the nodes, $\vec{r}_i$ is the
corresponding position in the embedding space, $\vec{n}_a$ is a normal to a
triangle $a$, and $\kappa$ is the bending rigidity.  The partition function is
given by the trace of the Boltzmann weight over all possible configurations of
the embedding variables $\{ \vec{r} \} $:
\begin{equation}
\label{e22}
{\cal Z} \;=\; \int [{\rm d}\vec{r}\,] \; \delta (\vec{r}_{cm}) \;
 \exp \left (- {\cal H}[\vec{r}\,] \, \right ).
\end{equation}
The center of mass of the membrane, $\vec{r}_{cm}$, is kept fixed to
eliminate the translational zero mode.  As there is no self-avoidance
term in the Hamiltonian, this model describes a phantom surface.

This model has been studied extensively with numerical methods
\cite{jan,wheater,espriu,uss}.  It has been found to have a
high-temperature crumpled (disordered) phase and a low-temperature
flat (ordered) phase, separated by a continuous phase transition ---
the {\it crumpling} transition. The behavior of the system in the flat
phase is governed by an infrared stable fixed point at $\kappa =
\infty$; the whole flat phase is critical.  The existence of an
ordered phase in a two-dimensional system with a continuous symmetry
and short range interactions is remarkable, given the Mermin-Wagner
theorem.  What stabilizes the flat phase are the out-of-plane
fluctuation of the membrane that couple to the in-plane phonon degrees
of freedom due to non-vanishing elastic moduli. Bending of the
membrane is necessarily accompanied by an internal stretching.  By
integrating out the phonon degrees of freedom, one is left with an
effective Hamiltonian with long-range interactions between the
Gaussian curvature fluctuations.

Most numerical simulations so far have used either local updating
methods, usually the Metropolis algorithm, or Fourier acceleration.
The Metropolis algorithm, apart from suffering from CSD, has very long
auto-correlations both in the flat phase and close to the crumpling
transition.  To establish this we have simulated the model
Eq.~(\ref{e22}), using the Metropolis algorithm, on a $L \times L$
square lattice, $L$ ranging from 8 to 64, and with periodic boundary
conditions.  We choose to simulate the model in the flat phase, at
$\kappa = 1.1$, were we know from previous simulations that the
auto-correlation time is indeed very long \cite{uss}.

To estimate the auto-correlations we measure the {\it square radius of
gyration}:
\begin{equation}
\label{e23}
 {\cal R}_g \;=\; 
 \left\langle \sum_i \vec{r}_i \cdot \vec{r}_i \right\rangle,
\end{equation}
i.e.\ the linear extent of the membrane in the embedding space.  This
is usually the ``slowest mode'' of the system.  From this we construct
the normalized auto-correlation function
\begin{equation}
\label{e24}
\rho(s) \;=\; 
 \frac{ \left <{\cal R}_g(t+s) {\cal R}_g(t) \right >
         - \left <{\cal R}_g \right >^2 }
      { \left <{\cal R}_g^2 \right > - 
         \left <{\cal R}_g^{} \right>^2 }, 
\end{equation}
and the integrated auto-correlation time (measured in units of sweeps)
\begin{equation}
\label{e25}
\tau \;=\; \frac{1}{2} + \sum_{s=1}^{\infty} \rho(s).
\end{equation}
This is shown in Table~1.  The errors on the
auto-correlation times are estimated from 10 to 20 independent runs, each few
hundred $\tau$ long.

\begin{table}
\begin{center}
\caption[ttab1]{{\small The integrated auto-correlation time 
 $\tau$ (in number of sweeps), together with the CPU-time per sweep $T_s$ (in
 ms). Column ${\it (a)}$ is for a lexicographic update of nodes, 
 while {\it (b)} is for random updating.  From a linear fit to 
 Eqs.~(\ref{e26}) and (\ref{e27}) we get the exponents $z_a$ 
 and $z_s$ and the corresponding amplitudes ${\cal A}_a$ and 
 ${\cal A}_s$.}}  \vspace{0.1in}
\begin{tabular}{|c|cc|cc|} \hline
    & \multicolumn{2}{c|}{\it (a)} & \multicolumn{2}{c|}{\it (b)} \\
$L$ & $\tau$ & $T_s$ 
      & $\tau$ & $T_s$ \\ \hline
8   &  219(15)     & 1.112       &  227(12)     & 1.036       \\
12  &  546(35)     & 2.784       &  567(30)     & 2.610       \\
16  &  1153(90)    & 4.900       &  1123(80)    & 4.688       \\
24  &  2714(150)   & 10.86       &  2534(220)   & 11.01       \\
32  &  4049(210)   & 22.23       &  4527(260)   & 21.68       \\
48  &  11443(600)  & 52.33       &  10347(550)  & 50.60       \\
64  &  20500(1200) & 97.59       &  19900(1300) & 98.18       \\ \hline
$z$ &  2.161(34)   & 2.143(23)   &  2.128(34)   & 2.188(31)   \\ 
${\cal A}$ 
    &  2.57(29)    & 0.0129(10)  &  2.83(28)   & 0.0111(13)   \\ \hline
\end{tabular}
\end{center}
\end{table}

As mentioned in the introduction, we also want to understand the
effect of randomness in the updating on performance.  Thus we have
repeated the simulations for two different updating schemes: the nodes
are sampled either at random or they are traversed in a {\it
lexicographic}\ order.  Lexicographic order means that the node at (intrinsic)
position $\vec{x}$ is always updated before $\vec{x}+\vec{e}_i
\;(i=1,2)$, except at the boundaries.  For a free field theory,
updated with the overrelaxation algorithm, only the latter scheme
reduces CSD \cite{tony}.

As an estimate of CSD we define the dynamical critical exponent $z_a$
using finite size scaling:
\begin{equation}
\label{e26}
 \tau \;\approx\; {\cal A}_a \xi^{z_a}.
\end{equation}
Here $\xi$ is some characteristic length scale of the system; in
our simulations, where the model is critical, this is the intrinsic
linear extent $L$.

To evaluate the performance of the algorithm, we also have to take
into account the amplitude ${\cal A}_a$ and how the computational work
performed per sweep, measured in CPU-time $T_s$, scales with system
size. This we include in Table~1.  Similarly to Eq.~(\ref{e26}) we
define an exponent $z_s$:
\begin{equation}
\label{e27}
 T_s \;\approx\; {\cal A}_s t^{z_s},
\end{equation}
where $t$ is now measured in ``real'' time (in ms).  Combining these
exponents and amplitudes, the performance --- the total ``cost'' of
the algorithm --- is given by ${\cal T}~=~{\cal A}_s {\cal A}_a
t^{z_a+z_s}$.  For the Metropolis algorithm we get:
\begin{equation}
{\cal T}_M \;=\; \left \{ 
 \begin{array}{ll}
   0.0332(45) \; L^{4.304(57)}  
     & \mbox{lexicographic updates,}  \\
   0.0314(48) \; L^{4.316(67)}  
     & \mbox{random updates.}
 \end{array}
 \right.
 \label{perf1}
\end{equation}
For this algorithm, the order in which nodes are updated is irrelevant.


\section{Hybrid overrelaxation}
Overrelaxation was introduced as a generalization of the heat-bath algorithm
for models with multi-quadratic actions \cite{adler}. In the original
formulation the new value of a field $\phi_i$ is chosen to be negatively correlated
with the old value.  Given a multi-quadratic action,
\begin{equation}
{\cal S} \;=\; \omega \left( \phi_i - {\cal F}_i[\phi_{j\neq i}] 
 \right )^2 \;+\; \{\mbox{ terms independent of $\phi_i$ }\},
\end{equation}
one chooses a local update of the field $\phi_i$ as
\begin{equation}
\phi_i \;\rightarrow \; 
 \phi_i^{\prime} \;=\;  
 (1-\zeta) \phi_i \;+\; \frac{\zeta }{\omega^2}{\cal F}_i \;+\;
 \frac{\sqrt{\zeta (2 - \zeta)}}{\omega} \xi \;. 
\end{equation}
$\xi$ is a Gaussian random variable of unit variance and $\zeta$
is a relaxation parameter. This update fulfills detail balance for $0
< \zeta \leq 2$; for $\zeta = 1$ it reduces to the standard heat-bath
algorithm, while for $\zeta = 2$ the field evolution becomes
deterministic and conserves energy (a micro-canonical simulation).  In
the latter case, in order to restore ergodicity, some amount of
standard ergodic updates have to be included.

This method has been applied successfully to variety of models; its
success based on it suppressing the usual random walk
behavior of local updating algorithms \cite{neal1}.  In order to
achieve the greatest reduction of CSD, both the relaxation parameter
$\zeta$, and the noise in the updating procedure, should be fine-tuned
\cite{tony,cucchi}.  Unfortunately, the usefulness of the method has
been limited by its restriction to multi-quadratic systems.

A number of generalizations of the overrelaxation have been proposed
\cite{genov}. They usually involve a transformation of the Boltzmann
distribution to the appropriate form Eq.~(9), and the introduction of
an accept-reject step to ensure detail balance.  This has the
disadvantage that the accept-reject step can enhance random walk
behavior by the algorithm and, in addition, the rejection probability
usually depends on some characteristics of the model and may not be
adjustable to a reasonable value.  This is nevertheless the approach
we will use.

We make a quadratic approximation to the Hamiltonian Eq.~(\ref{e21})
and then apply hybrid overrelaxation (with $\zeta =2$).  We treat the
non-linear bending energy term by assuming that normalization in the
denominator is {\it constant} for all the triangles, i.e.\ we write
the normals as
\begin{equation}
\vec{n}_a (\vec{r}) \;=\;
 \frac
   {(\vec{r}_i - \vec{r}_j) \times (\vec{r}_i - \vec{r}_{k})}
   { \sqrt{\left [(\vec{r}_i - \vec{r}_i) \times 
 (\vec{r}_i - \vec{r}_{k}) \right ]^2} }
\; \approx \; \frac{ \vec{r}_j \times \vec{r}_{k}
                - \vec{r}_i \times (\vec{r}_j + \vec{r}_{k}) }
              {\Lambda} \;'
\end{equation}
were $\{i,j,k\}$ are the nodes defining triangle $a$.  
Then, a quadratic approximation to the Hamiltonian can be written as 
\begin{equation}
 \label{Happrox}
{\cal H}_A = \sum_{\langle ij\rangle} \left (\vec{r}_i - \vec{r}_j \right )^2
   - \frac{\kappa}{\Lambda^2} \sum_{\langle ab \rangle} 
 \left [ \vec{r}_j \times \vec{r}_k
   - \vec{r}_i \times (\vec{r}_j + \vec{r}_k) \right] \cdot
   \left[ \vec{r}_k \times \vec{r}_l - \vec{r}_i \times 
(\vec{r}_k + \vec{r}_l)\right ],
\end{equation}
where the triangles $a=(i,j,k)$ and $b=(i,k,l)$ are adjacent.  Since the
approximate Hamiltonian is quadratic in $\vec{r}_i$, we can write:
\begin{equation}
  \label{hamri}
{\cal H}_A = \vec{r}_i \cdot (\hat{M} \vec{r}_i)
 \;+\;  \vec{C} \cdot \vec{r}_i  
 \;+\; \{ \mbox{terms independent of } \vec{r}_i \} .
\end{equation}
The matrix $\hat{M}$ and the vector $\vec{C}$ are easily computed:
\begin{eqnarray}
\hat{M}_{aa} &=& 6 - \frac{\kappa}{\Lambda^2} \sum_{j=1}^6 \sum_{b \neq a}
\left( r^{\, (b)}_{j} - r^{\, (b)}_{j-1} \right)
\left( r^{\, (b)}_{j+1} - r^{\, (b)}_{j} \right) \;,\nonumber \\
\hat{M}_{a \neq b} & = & \frac{\kappa}{\Lambda^2} \sum_{j=1}^6
\left( r^{\, (a)}_{j} - r^{\,
(a)}_{j-1} \right) \left( r^{\, (b)}_{j+1} - r^{\, (b)}_{j} \right)\;,
\label{matM}
\end{eqnarray}
and
\begin{eqnarray}
C^{(a)} & = & - \sum_{j=1}^6 \left\{ r_j^{(a)} \left[ 
  \frac{\kappa}{\Lambda^2} \left( 
  \vec{r}_{j} \cdot (2 \vec{r}_{j-1} + 2 \vec{r}_{j+1} 
   - \vec{q}_{j} -\vec{q}_{j-1} ) 
  - (\vec{r}_{j+1} + \vec{r}_{j-1})^2\right. \right. \right. \nonumber \\
  & & \left. \left. \left. \mbox{} + \vec{r}_{j-1} \cdot \vec{r}_{j-2} 
  + \vec{r}_{j+1} \cdot \vec{r}^{}_{j+2} 
  \right) - 2 \frac{}{} \right] 
  + \frac{\kappa}{\Lambda^2} q^{(a)}_{j} \left( 
  \vec{r}_{j+1} - \vec{r}_{j} \right)^2 \right\} \;.
\label{vecC}
\end{eqnarray}
The indices $a$ and $b$ label the component of the
fields in the embedding space, and the index $j$ labels the neighbors
of node $i$, including its next-to-nearest neighbors $\vec{q}_j$, in a
cyclic manner.

The constant energy surface, ${\cal H}_A(\vec{r}_i) = k$, is a
multi-quadratic function --- in our case an ellipsoid in the embedding
space.  To find the new (overrelaxed) embedding position
$\vec{r}_i^{\, \prime}$, we can diagonalize the matrix $\hat{M}$ and
apply overrelaxation to the transformed variables.  This involves some
amount of calculation; a quicker and sufficient method is to apply
overrelaxation in a random sequence to each of the embedding
positions $r_i^{(a)}$, $a=1,2,3$.  Once the trial position has
been chosen, it is accepted or rejected according to a Metropolis
test.

\begin{figure}
\epsfxsize=4.5in \centerline{ \epsfbox{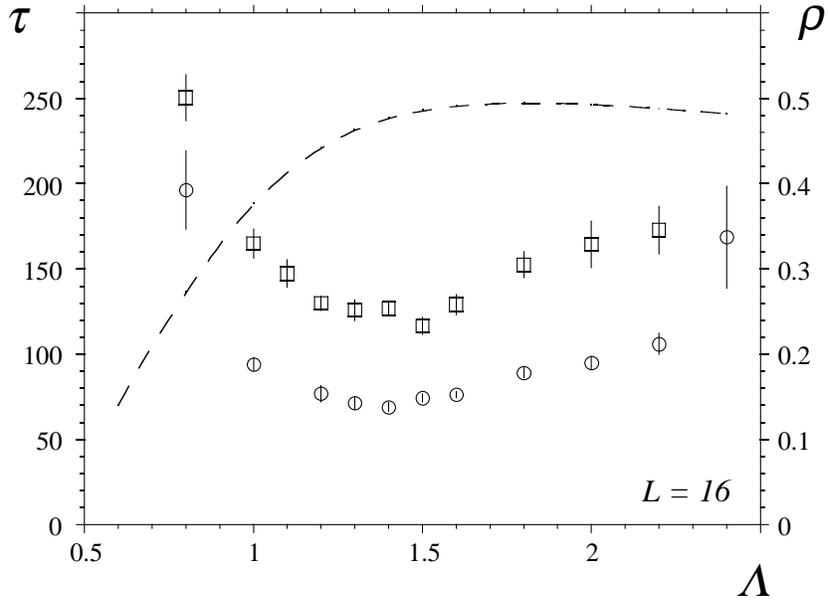}}
\caption[fig1]{\small The integrated auto-correlation time $\tau$,
{\it vs}.\ the ``normalization'' parameter $\Lambda$, for the hybrid
overrelaxation algorithm.  This is shown both for a random (squares)
and a lexicographic (circles) updating.  The lattice size is $16^2$.
The dashed line is the corresponding acceptance rate $\rho$ in the
Metropolis test.}
\end{figure}

An important feature of this approximation is the parameter $\Lambda$.
Although introduced as a substitution for the normalization of the
normals, it can be tuned to optimize the performance of the
algorithm by minimizing the rejection probability in the Metropolis
test.

We have applied this method in the flat phase ($\kappa = 1.1$) for
both random and lexicographic updating.  To ensure ergodicity we also
include a random amount of standard Metropolis updates (about $20\%
$).  In Fig.~1 we show the integrated autocorrelation time {\it vs}.\
$\Lambda$, for a lattice size $L = 16$. We also include the
corresponding acceptance rate in the Metropolis test.  In this case,
contrary to the Metropolis algorithm, lexicographic updating reduces
$\tau$ by about $30\%$, independent of system size.  More important,
for a suitable choice of $\Lambda$, $\tau$ is reduced by a factor of
15 relative to the Metropolis algorithm.  As it might be expected, the
optimal choice of $\Lambda$ corresponds to maximizing acceptance rate
in the Metropolis test. Surprisingly this value is much higher than
one would expect from the average length of the un-normalized normals
($\langle |\vec{n}_{un}| \rangle \approx 0.3$ for $\kappa = 1.1$).
This implies that, in order to enhance the acceptance rate, it is
convenient to suppress the bending energy term in the approximation.

\begin{table}
\begin{center}
\caption[ttab2]{{\small  Same as Table 1, except the algorithm used is the 
 approximate hybrid overrelaxation.  Again {\it (a)} corresponds
 to lexicographic and {\it (b)} to random updating.  
 The optimal value of the parameter $\Lambda$ is also included.}}  
 \vspace{0.1in}
\begin{tabular}{|c|ccc|ccc|} \hline
    & \multicolumn{3}{c|}{\it (a)} & \multicolumn{3}{c|}{\it (b)} \\ 
$L$ & $\Lambda$ &$\tau$ & $T_{sweep}$        
    & $\Lambda$ &$\tau$ & $T_{sweep}$      \\ \hline
8   & 1.08   &  18.5(7)        & 1.247      
    & 1.08   &  31.9(1.2)      & 1.136            \\
12  & 1.23   &  36.6(1.1)      & 3.078      
    & 1.22   &  66.9(2.0)      & 2.766            \\
16  & 1.38   &  70.6(8.0)      & 5.735      
    & 1.35   &  113.3(6.0)     & 5.122            \\
24  & 1.57   &  150(11)        & 13.42      
    & 1.58   &  279(18)        & 11.83            \\
32  & 1.92   &  269(25)        & 27.27      
    & 1.94   &  484(42)        & 24.12            \\
48  & 2.68   &  640(40)        & 62.69      
    & 2.60   &  1096(90)       & 53.19            \\
64  & 3.09   &  1120(105)      & 118.7      
    & 3.20   &  2600(180)      & 103.5            \\ \hline
$z$ &        &  1.990(31)      & 2.190(14)                 
    &        &  2.065(30)      & 2.163(18)        \\  
${\cal A}$&  &  0.275(24)      & 0.0132(6)     
    &        &  0.405(34)      & 0.0127(8)       \\ \hline
\end{tabular}
\end{center}
\end{table}

We have repeated this analysis for lattices sizes $L = 8$ to 64.  In
Table~2 we show the optimal values of $\Lambda$, the corresponding
integrated auto-correlation time and the CPU-time for a sweep.  From
this data we extract, as before, the exponents $z_a$ and $z_s$ and the
amplitudes ${\cal A}_a$ and ${\cal A}_s$, and obtain the following
performance:
\begin{equation}
{\cal T}_O \;=\; \left \{
 \begin{array}{ll}
   0.00364(36) \; L^{4.167(46)}
     & \mbox{lexicographic updates,}  \\
   0.00514(30) \; L^{4.228(48)}
     & \mbox{random updates.}
 \end{array}
 \right.
 \label{perf2}
\end{equation}
Although hybrid overrelaxation does not reduce CSD, it
gives an improvement of one order of magnitude over the Metropolis
algorithm, provided the nodes are updated in a lexicographic order and
the ``normalization'' parameter $\Lambda$ is properly adjusted.


\section{A unigrid Monte Carlo algorithm}
The critical slowing down of traditional Monte Carlo algorithms arises
mainly from the fact that the update is local, and thus the system
takes a random walk through the configuration space.  This can be
improved by using collective mode (non-local) updating such as
multigrid methods \cite{orgmult,multigrid}.  The basic idea is to
consider a sequence of coarser lattices (levels) in addition to the
original lattice.  At each level the system is updated using
traditional methods but, as this is repeated recursively at all length
scales, the long wavelength modes are equilibrated much faster.

There are several basic ingredients to a multigrid algorithm: a {\it
restriction} operator and the corresponding {\it interpolation}
operator, or kernel, are needed to map the system onto the coarser
lattices and back; an updating algorithm, such as Metropolis, is
applied at each level;  finally, one has to choose how to traverse
the different levels.

For a crystalline membrane it is not possible to construct an exact
interpolation operator between different levels due to the complexity
of the Hamiltonian Eq.~(\ref{e21}).  This problem can be circumvented
by using an alternative implementation, the {\it unigrid} method, in
which the coarse lattices are simply defined as subdivisions of the
original one; the update of a coarse lattice acts on blocks of the
original fields.  For the update, the choice is usually between a
piecewise constant or a piecewise linear kernel.  A piecewise kernel
simply shifts all of the fields in the block by a uniform value.  A
piecewise linear kernel shifts the fields by a value linearly
interpolated between zero, at the boundary, and a maximum value at the
center of the block.  The shift operation is one of the global symmetries of
the system.

Several considerations should be made in choosing a kernel.
The piecewise linear kernel has the advantage that the acceptance rate of
the proposed moves does not depend on the block size. For a crystalline
membrane, on the other hand, this is outweighed by the computational
cost which, as all the normal-normal interactions in the block have to
be recalculated, grows linearly with the block size.  Hence a
piecewise constant kernel is preferable, since the sole contribution
to the energy change comes from the boundary.

We parameterize a non-local change of the configuration, when we
shift a block $\Lambda_k$ at level $k$, as:
\begin{equation}
 \label{unshift}
  \vec{r}^{\; \prime}_i  \;= \; \left\{  
 \begin{array}{ll}
  \vec{r}_i + \epsilon_k \vec{x} & \;\;\;\;
    \mbox{if}\;\;  i \in \Lambda_k\\
  \vec{r}_i  & \;\;\;\; \mbox{otherwise,} 
 \end{array}\right.
\end{equation}
where $\vec{x}$ is some normalized random noise and $\epsilon_k$ is the
amplitude of the shift.  A necessary prerequisite for the unigrid method to be
efficient is that the energy cost does not grow too fast with the perimeter
$L_k$ of the block.  Stated differently: in order to maintain a constant
acceptance rate in the Metropolis test, the amplitudes have to be scaled like
\begin{equation}
 \label{alphadef}
 \epsilon_k \sim L_k^{-\alpha}.
\end{equation}
If the exponent $\alpha$ is too big, i.e.\ of order unity, it is unlikely that
any multigrid algorithm will reduce CSD.

Following the analysis of \cite{GP}, we can estimate whether the unigrid update
has a chance of improving the dynamical behavior in the case of
a crystalline membranes.
Assuming that the probability distribution is nearly Gaussian, one
can approximate the acceptance rate by
\begin{equation}
  \label{accept}
  \Omega(\epsilon) \sim \mbox{erfc}(\sqrt{h}/2),
\end{equation}
where $h = \langle \Delta {\cal H}\rangle$ is the average change in 
energy and $\epsilon$ is the amplitude of the proposed move. 
For a piecewise constant kernel we take the Hamiltonian
Eq.~(1), insert Eq.~(\ref{unshift}) and expand the change in the 
energy in powers of $\epsilon$:
\begin{equation}
 \langle \Delta {\cal H}(\{\vec{r}_i\};\vec{x}) \rangle \;=\;
 \epsilon \;  \langle F_1(\{\vec{r}_i\};\vec{x}) \rangle \; + \;
 \epsilon^2 \; \langle F_2(\{\vec{r}_i\};\vec{x}) \rangle \; + \;
 {\cal O}(\epsilon^3).
 \label{cheng}
\end{equation}
The key observation is that under a global sign change, $\vec{r}_i
\rightarrow -\vec{r}_i \; \forall i$, the function $F_1$ changes sign
(as each term is a product of {\it odd} number of fields $\vec{r}_i$),
hence $\langle F_1(\{\vec{r}_i\};\vec{x}) \rangle = 0$.  The leading
contribution to $h$ is therefore proportional to $\epsilon^2$.  At the
same time, the number of terms contributing to Eq.~(\ref{cheng})
depends linearly on the length of the boundary.  Therefore, in order
to maintain a constant acceptance rate, one should scale the
amplitudes as $\epsilon \sim L_B^{-1/2}$.  This agrees with our
numerical simulations where we find $\alpha = 0.52(1)$.

Another free parameter in the unigrid method is the relative frequency
with which different levels are updated.  One must strike a balance
between the effectiveness of block moves and their relative
computational cost.  Two general schemes are used in the literature;
the $V$ and $W$--cycles.  In a $V$--cycle the levels are updated
consecutively, from the finest to the coarsest and back, whereas a
$W$--cycle recursively applies a $V$--cycle at each visited level,
spending more time in updating coarser levels.  For multigrid
algorithms, where the lattices size decreases between levels, a
$W$--cycle is preferable, provided that the interpolations
between levels is not too time consuming.  For a unigrid algorithm the
computational cost increases with the block size, as discussed above,
and, depending on the exponent $\alpha$, a $V$ or $W$--cycle will be
advantageous.  For a piecewise constant kernel the computational cost
scales like $L_B$ and a $W$-cycle might be advantageous, while for a
piecewise linear kernel the computational cost scales linearly with
the area of the block and a $V$--cycle would be better.

\begin{table}
\begin{center}
\caption[ttab3]{{\small The auto-correlation and
 CPU-times for the unigrid algorithm.  Results are shown both for  
 $V$ and $W$--cycles. }}  
\vspace{0.1in}
\begin{tabular}{|c|cc|cc|} \hline
    & \multicolumn{2}{c|}{\it $V$-cycle} & 
      \multicolumn{2}{c|}{\it $W$-cycle} \\
$L$ & $\tau$ & $T_s$ & $\tau$ & $T_s$ \\ \hline
8   & 17.1(0.4)    & 3.123       & 14.0(0.3)    & 4.195      \\
16  & 45.8(3.2)    & 16.325      & 30.1(0.9)    & 25.904     \\
32  & 115.2(6.1)   & 107.18      & 51.0(2.1)    & 203.24     \\
64  & 269.3(19.1)  & 524.70      & 96.8(8.7)    & 1324,25    \\ \hline
$z$ & 1.349(28)    & 2.489(55)   & 0.955(27)    & 2.788(45)  \\
${\cal A}$
    & 1.040(77)    & 0.0175(31)  & 3.96(14)     & 0.0123(18) \\ \hline
\end{tabular}
\end{center}
\end{table}

As before, we have simulated a crystalline membrane in the flat phase
and for lattice sizes $L = 8, 16, 32$ and 64, using the unigrid
algorithm and updating the system at each level with the Metropolis
algorithm.  We repeated the simulation both for a $V$ and $W$--cycle.
In Table~3 the show the measured value of the auto-correlation and
CPU-times, from which we determine the overall performance of the
algorithm:
\begin{equation}
{\cal T}_U \;=\; \left \{
 \begin{array}{ll}
   0.0182(35) \; L^{3.838(84)}  & \mbox{$V$--cycle}, \\
   0.0242(39) \; L^{3.743(72)}  & \mbox{$W$--cycle}.
 \end{array}
 \right.
 \label{perf3}
\end{equation}
In both cases the unigrid algorithm reduces CSD, albeit not very much,
but enough to make it worthwhile for large membranes.  
For the $W$--cycle this implies a dynamical CSD exponent 
$z \approx 1.7$, although this value is probably to large,
as we observe strong finite size effects in the fit to Eq.~(\ref{e26});
if we exclude the smallest volume ($L = 8$), we get 
$z \approx 1.6$.  In conclusion, although the  $W$--cycle
is more time consuming per sweep, its performance is      
better than $V$--cycle in simulations of 
membranes of size $L \gsim 20$.

\section{Discussion}

Comparing the performance of these different algorithms, Eqs.\
(\ref{perf1}), (\ref{perf2}), and (\ref{perf3}), we see that, in the
simulation of crystalline membranes on realistic lattice sizes
($L$~=~10 to 200), both the hybrid overrelaxation and the unigrid
algorithm reduce the computational cost by an order of magnitude over
the Metropolis algorithm.  As the unigrid algorithm also reduces the
dynamical exponent $z$, especially using a $W$--cycle, it is clearly
the best choice for large membranes.  In the particular case we
studied in this paper, large means $L \gsim 50$, although this value
may depend on the simulation parameters (i.e.\ $\kappa$).

We would also like to emphasize that, in order to achive optimal
performance of the hybrid overrelaxation, it is imperitive to adjust the
noise in the updating procedure (to use lexicographic updating), and to
tune the ``normalization'' parameter $\Lambda$ appropriatly.

An alternative algorithm used to simulate crystalline membranes is a
combination of Langevin updates with Fourier acceleration
\cite{wheater,espriu}.  This algorithm is known to substantially
reduce CSD, although the gain is lost to some extent in the large
computational overhead.  This method is also complicated by systematic
errors induced by using a finite time step $\Delta t$; this
necessitates an extrapoltion to $\Delta t = 0$, which can itself
become time consuming.  Nevertheless, it would be interesting to know
how well this algorithm performes, in realistic simulations, compared
to the algorithms we have studied in this paper. Unfortunately, we do
not have an estimate of its performance in similar conditions (e.g.,
using the same computers) for comparison.

An obvious extension of the methods studied in this paper, is to
combine hybrid overrelaxation with the unigrid algorithm.  It is
possible to maximize the shift of a block in a unigrid update, by
choosing it in a deterministic and energy preserving way, improving
the performance even further.  This is though more complicated to
implement and it is under investigation.

There are few application in which the hybrid overrelaxation might be
more advantageous compared to the unigrid algorithm.  
Overrelaxation can be parallelized in a
straightforward manner, although it could be difficult to define a
lexicographic update in that case.  
It also easy to adapt hybrid overrelaxation to
modified versions of the model like, for example, {\it self-avoiding}
crystalline membranes, which are of great physical interest.  In that
case, a proposed updated is rejected if it leads to self-intersection
of the membrane.  Intuitively, a non-local change, like one proposed
by the unigrid algorithm, is more likely to be rejected --- hybrid
overrelaxation might turn out to be more effective.

Finally, we would like to point out that both these updating
algorithms can be used in simulations of {\it fluid} membranes with
extrinsic curvature \cite{fluid}.  In that case, the surface
fluctuates in the embedding space, and its connectivity matrix changes
dynamically.  It is known that simulations of fluid membranes suffer
from even longer auto-correlation times than their crystalline
counterparts.  For fluid membranes, the random nature of the lattice
causes some complications in implementing hybrid overrelaxation and
unigrid algorithms.  For example, it is not obvious how to define lexicographic
ordering.  A possible method would be to propagate the
updates outwards from a randomly chosen node, i.e.\ by traversing the
lattice in steps of increasing geodesic distance from a marked point.
For the unigrid algorithm the
problem is that a random lattice cannot be divided into regular
blocks.  Again, blocks could be defined as nodes within a given
geodesic distance from some randomly chosen node.

\section*{Acknowledgements}
MF wishes to thank Bengt Petersson for his hospitality at Bielefeld,
where part of this work was carried out, and Alan Sokal for
stimulating discussions.  We are grateful to NPAC (North-East Parallel
Architecture Center) for the use of their computational facilities.
The research of GT was supported by the Alexander von Humboldt
Stiftung and the Deutsche Forschungsgemeinschaft.  The research of MF
was supported by the Department of Energy U.S.A.\ under contract
No.~DE-FG02-85ER40237 and by a Syracuse University Graduate
Fellowship.


\begin{thebibliography}{99}

\raggedright

\bibitem{adler}
 S.L.~Adler, Phys.~Rev.\ {\bf D23} (1981) 2901.

\bibitem{lan}
 G.G.~Batrouni, G.R.~Katz, A.S.~Kronfeld, G.P.~Lepage, B.~Svetitsky,
  and K.G.~WIlson, Phys.~Rev.\ {\bf D32} (1985) 2736.

\bibitem{orgmult}
 G.~Parisi, in {\it Progress in Gauge Field Theory}, (Charges 1983),
  G.~t'Hooft {\it et.~al.\ } (eds.), Plenum, New York, 1984; \\
 J.~Goodman and A.D.~Sokal, Phys.~Rev.~Lett.\ 56 (1986) 1015;
  Phys.~Rev.\ {\bf D40} (1989) 2035.

\bibitem{swendsen}
 R.H.~Swendsen and J.-S.~Wang, Phys.~Rev.~Lett.\ {\bf 58} (1987) 86.

\bibitem{polyakov}
 A.~Polyakov, Nucl.~Phys.\ {\bf B268} (1986) 406.

\bibitem{jan}
 J.~Ambj{\o}rn, B.~Durhuus and T.~Jonsson, Nucl.~Phys.\
  {\bf B316} (1989) 526.

\bibitem{wheater}
 R.G.~Harnish and J.F.~Wheater, Nucl.~Phys.\ {\bf B350} (1991) 861; \\
 J.F.~Wheater, Nucl.~Phys.\ {\bf B458} (1996) 671  
  ({\tt hep-lat/9503008}).

\bibitem{espriu}
 M.~Baig, D.~Espriu and A.~Travesset, Nucl.~Phys.\ {\bf B426} (1994)
  575;\\
 D.~Espriu and A.~Travesset, Phys.~Lett.\ {\bf B356} (1995) 329
  ({\tt hep-lat/9505018}); Nucl.~Phys.\ {\bf B47} (Proc.\ Suppl.) 
  (1996) 637 ({\tt hep-lat/9509062}).

\bibitem{uss}
M.J.~Bowick, S.M.~Catterall, M.~Falcioni, G.~Thorleifsson and
 K.N.~Anagnostopoulos, Nucl.~Phys.\ {\bf B47} (Proc.~Suppl.) (1996) 838
 ({\tt hep-lat/9509074});  J.~Phys.~I France {\bf 6} (1996) 1321
 ({\tt  cond-mat/9603157});  Nucl.~Phys. {\bf B53} (Proc.~Suppl.)
 (1997) 746 ({\tt  hep-lat/9608044}); \\
M.~Falcioni, M.J.~Bowick, E.~Guitter and G.~Thorleifsson,
 Europhys.~Lett.\ {\bf 38} (1997) 67 ({\tt cond-mat/9610007}).

\bibitem{tony}
I.~Horv\'{a}th and A.D.~Kennedy, {\it The Local Hybrid Monte Carlo
 Algorithm for Free Field Theory: Reexmaining Overrelaxation},
 FSU-SCRI-97-94 ({\tt hep-lat/9708024}).

\bibitem{neal1}
 R.M.~Neal, {\it Suppressing Random Walks in Markov Chain Monte Carlo
  Using Ordered Overrelaxation}, Technical report No.~9508, Dept.\ of
  Statistics, University of Toronto ({\tt bayes-an/9506004}).

\bibitem{genov}
 F.R.~Brown and T.J.~Woch, Phys.~Rev.~Lett.\ {\bf 58} (1987) 2394; \\
 M.~Creutz, Phys.~Rev.\ {\bf D36} (1987) 515; \\
 P.J.~Green and X.~Han, in {\it Stocastic Models, Statistical Methods, 
  and Algorithms in Image Analysis}, P.~Barone {\it et.~al.\ } 
  (eds.), Lecture Notes in Statistics, Berlin (1992), Springer-Verlag; \\
 Z.~Fodor and K.~Jansen, Phys.~Lett.\ {\bf B331} (1994) 119
  ({\tt hep-lat/9403024});

\bibitem{multigrid}
 W.L.~Briggs, {\em A Multigrid Tutorial}, SIAM, Philadelphia, 1987; \\
 W.~Hackbush, {\em Multigrid Methods and Applications}, Springer, Berlin,
  1985.

\bibitem{cucchi}
 A.~Cucchieri and T.~Mendes, Nucl.~Phys.\ {\bf B471} (1996) 263
  ({\tt hep-lat/9511020}).

\bibitem{GP} 
 M.~Grabstein and K.~Pinn, Phys.~Rev.~D {\bf 45}(12) (1992) 4372
  ({\tt hep-lat/9204016}).

\bibitem{fluid}
 J.~Ambj\o rn, A.~Irback, J.~Jurkiewicz, B.~Petersson, Nucl.~Phys.\
  {\bf B393} (1993) 571 ({\tt hep-lat/9207008}); \\
 M.~Bowick, P.~Coddington, L.~Han and G.~Harris, Nucl.~Phys.\ 
  {\bf B394} (1993) 791 ({\tt hep-lat/9209020});  \\
 K.~Anagnostopoulos, M.~Bowick, P.~Coddington, M.~Falcioni, L.~Han,
  G.~Harris and E.~Marinari, Phys.~Lett.\ {\bf B317} (1993) 102
  ({\tt hep-th/9308091}). 
\end{thebibliography}
\end{document}